# Influence of Pulse width and Rabi frequency on the Population dynamics of three-level system in two-photon absorption process

Nam-Chol Kim, [†,*] Myong-Chol Ko,[†] Song-Jin Im,[†] Zhong-Hua Hao [‡]

[†] Department of Physics, Kim Il Sung University, Pyongyang, D. P. R. Korea

[‡] Department of Physics, Key Laboratory of Acoustic and Photonic Materials and Devices of Ministry of Education, Wuhan University, Wuhan 430072, P. R. China

**Abstract :** We investigate the population dynamics of the three-level system in the two-photon absorption (TPA) process, mainly focusing the influence of pulse width and Rabi frequency on the population dynamics of the system. We observe the dependency of the population with the Rabi frequency and the pulse width. We also show that the arbitrary superposition state consisted in two states, upper state and lower state, is possible by controlling the pulse width and Rabi frequency. The results obtained can be used to the case of more complex multilevel system and they can be valuable for coherent quantum control in quantum information processing.

*Email address: elib.rns@hotmail.com

## 1. Introduction

Two-photon absorption (TPA) in nanostructures has been the focus of attention of many researchers because of their potential application in a wide range of optical devices[1-3]. TPA is one of the most basic radiation-matter interaction mechanics. It consists in the excitation of an atom or molecule from a lower quantum state(ground state) |1> to an upper quantum state (excited state) |2> of the same parity as |1> in a single step. In this case the initial and final states cannot be connected through an electric-dipole transition. Thus parity conservation implies that two light quanta must be absorbed simultaneously. The theory of TPA was first developed by Maria Göppert-Mayer in 1931[4]. As a multiphoton process, TPA is closely related to Raman scattering. In the latter process, one photon is absorbed while the other is simultaneously emitted, the energy difference being retained by the molecule. While spontaneous Raman scattering was observed as early as 1928[5], TPA was not observed until 1961[6] after the advent of the laser. The reason for that delay in the observation of the two multiphoton processes lies in the fact that while in spontaneous Raman scattering the scattered light intensity is proportional to the intensity of the incoming radiation, in TPA the power absorbed is proportional to the square of the intensity of the incoming field and thus higher excitation energy is required for TPA.

One of the many applications of TPA is in spectroscopy. The use of two-photon absorption for spectroscopy studies has advantages over the single-photon process because of the restriction on selection rules; e.g., when a single photon absorption is forbidden by selection rules, a higher-order absorption process may be allowed [7]. Another advantage of two-photon absorption spectroscopy is the ability to study the properties truly characteristic of the crystalline volume because of the small values of the nonlinear absorption coefficient [8].

Nowadays, coherent control of quantum states with light has attracted great interest as a means to influence the outcome of a quantum-mechanical interaction. In principle, the quantum system can be controlled towards a desired state by its interaction with light [9]. Steering quantum processes in atoms and molecules through the manipulation of the properties of optical fields is the goal of coherent quantum control [10-12]. In particular, control of population transfer in multiphoton transitions with laser pulses has been achieved through several methods. A common strategy for ultrashort pulses is the manipulation of the spectral phases and amplitudes of the frequency components of the fields exciting the medium, resulting in pulses or sequences of pulses with very diverse temporal shapes which enhance or suppress the coupling between selected states by exploiting quantum interference effects [13–19]. Other techniques using longer pulses such as adiabatic methods [20–22] and π -pulse polychromatic control [23-25] can be implemented to induce complete population transfer between a pair of quantum states. Multiphoton control techniques are valuable in several areas, including nonlinear spectroscopy [26], in femtochemistry and biology [27], or in quantum information and in the quantum engineering of light states [28].

In recent work [29] it was shown analytically and numerically that the population of the quantum system can be controlled by changing the duration, shape and intensity of the input pulse for a TPA process. Recently, it was also shown that the two-photon transition rate of quantum dots coupled to nanocavities are enhanced by up to several orders of magnitude relative to quantum dots in bulk host and proposed a simple cavity design to achieve QD degenerate two-photon absorption with low power, which allows us to employ enhanced TPA to coherently excite QDs and enable generation of indistinguishable single photons on demand [30]. So far, most reported multi-photon studies on quantum systems were focused on two-photon absorption and excitation processes with changing of such parameters as pulse shape, pulse chirp, pulse area, detuning and the size of quantum systems like quantum dots(QDs).

In this paper, we consider the population dynamics of a three-level system mainly focusing the influence of pulse width and the Rabi frequency to the occupation probabilities of the quantum system.

**2. Theoretical Model**

Now, we are considering the two-photon absorption in a three-level system (e.g., semiconductor quantum dot; QD), the schematic of which is shown in Fig. 1.

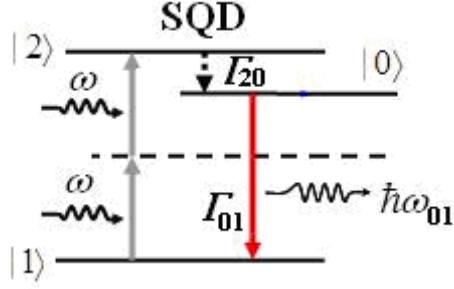

Fig. 1. Energy diagram of two-photon three-level model.

We start from the Hamiltonian in the Schrödinger picture. In the Schrödinger picture, the Hamiltonian of the QD system can be written in the form of the sum of the unperturbed part and interaction part as follow,

$$\hat{H}_s = \hat{H}_0 + \hat{H}_1. \tag{1}$$

The unperturbed Hamiltonian of the QD three-level system is given as follow,

$$\hat{H}_0 = \hbar\omega_{20}|2\rangle\langle 2| + \hbar\omega_{01}|1\rangle\langle 1|, \tag{2}$$

where $\omega_{ij} = (E_i - E_j)/\hbar$ are the angular frequencies of the transitions, with $E_i$ being the energies of the different quantum state, the intermediate state $|0\rangle$ of which is taken to be the origin of energies. $\omega$ is the angular frequency of the incoming laser field and $\Gamma_{20}$, $\Gamma_{01}$ are the decay rate of the states $|2\rangle$ and $|0\rangle$, respectively. The allowed electric-dipole transitions are $|1\rangle \leftrightarrow |0\rangle$ and $|0\rangle \leftrightarrow |2\rangle$, because levels $|2\rangle$ (excited state) and $|1\rangle$ (ground state) have the same parity, and contrary to that of intermidiate state $|0\rangle$. Thus the dipole-moment operator can be written as

$$\boldsymbol{\mu} = \mu_{20}|2\rangle\langle 0| + \mu_{02}|0\rangle\langle 2| + \mu_{01}|0\rangle\langle 1| + \mu_{10}|1\rangle\langle 0|, \tag{3}$$

where $\mu_{ij} = \langle i|\mu|j\rangle$, that can be taken to be real without loss of generality through proper choice of the basis states phases $(\mu_{ij} = \mu_{ji})$. The interaction Hamiltonian of an atom located at $z = 0$ reads

$$\hat{H}_1 = -\boldsymbol{\mu} \cdot \boldsymbol{E}(t), \tag{4}$$

where we consider a classical monochromatic electromagnetic field of the form

$$\boldsymbol{E}(t) = \boldsymbol{e}E(t). \tag{5}$$

In order to transform the Hamiltonian from the Schrödinger picture to the interaction picture, we introduce a unitary operator as

$$\hat{U}(t) = e^{-i\hat{H}_0 t/\hbar} = e^{-i\omega_{20}t}|2\rangle\langle 2| + |0\rangle\langle 0| + e^{i\omega_{01}t}|1\rangle\langle 1|. \quad (6)$$

The interaction Hamiltonian in the interaction picture is defined by

$$\hat{H}_I = \hat{U}^{-1}\hat{H}_1\hat{U}. \quad (7)$$

The interaction Hamiltonian in the Schrödinger picture is as follow,

$$\hat{H}_1 = -\boldsymbol{\mu} \cdot \boldsymbol{E}(t) = -\hbar\left[\Omega_2(t)|2\rangle\langle 0| + \Omega_2(t)|0\rangle\langle 2| + \Omega_1(t)|0\rangle\langle 1| + \Omega_1(t)|1\rangle\langle 0|\right], \quad (8)$$

where $\Omega_2(t) = \dfrac{(\boldsymbol{\mu}_{20}\cdot\boldsymbol{e})E(t)}{\hbar}$ and $\Omega_1(t) = \dfrac{(\boldsymbol{\mu}_{01}\cdot\boldsymbol{e})E(t)}{\hbar}$ are the Rabi frequencies of the transitions $|2\rangle \leftrightarrow |0\rangle$ and $|0\rangle \leftrightarrow |1\rangle$, respectively.

The interaction Hamiltonian in the interaction picture is given by

$$\hat{H}_I = \hat{U}^{-1}\hat{H}_1\hat{U} = -\hbar\left[\hat{\sigma}_{20}^{(i)}\Omega_2(t) + \hat{\sigma}_{01}^{(i)}\Omega_1(t) + H.c.\right] \quad (9)$$

where $\hat{\sigma}_{20}^{(i)} = e^{i\omega_{20}t}|2\rangle\langle 0|$ and $\hat{\sigma}_{01}^{(i)} = e^{i\omega_{01}t}|0\rangle\langle 1|$ are the dipole operators of the transition in the interaction picture. The population dynamics of the system is described by the master equation in the Lindblad form

$$\frac{d}{dt}\rho = -\frac{i}{\hbar}[\hat{H}_I, \rho] + \hat{L}(\rho), \quad (10)$$

where the dissipative term $L(\hat{\rho})$ is

$$\hat{L}(\rho) = \frac{\Gamma_{20}}{2}(2\hat{\sigma}_{02}\rho\hat{\sigma}_{20} - \hat{\sigma}_{22}\rho - \rho\hat{\sigma}_{22}) + \frac{\Gamma_{01}}{2}(2\hat{\sigma}_{10}\rho\hat{\sigma}_{01} - \hat{\sigma}_{00}\rho - \rho\hat{\sigma}_{00}) \quad (11)$$

Combining Eqs. (10) - (11), the optical Bloch equations for the population of the system are obtained:

$$\dot{\rho}_{11} = i\Omega_1(t)[\rho_{01} - \rho_{10}] + \Gamma_{01}\rho_{00}, \quad (12b)$$

$$\dot{\rho}_{22} = i\Omega_2(t)(\rho_{02} - \rho_{20}) - \Gamma_{20}\rho_{22}, \quad (12c)$$

$$\dot{\rho}_{01} = -i\Omega_1(t)(\rho_{00} - \rho_{11}) + i\Omega_2(t)\rho_{21} - i\omega_{01}\rho_{01} - \frac{\Gamma_{01}}{2}\rho_{01}, \quad (12d)$$

$$\dot{\rho}_{10} = i\Omega_1(t)(\rho_{00} - \rho_{11}) - i\Omega_2(t)\rho_{12} + i\omega_{01}\rho_{10} - \frac{\Gamma_{01}}{2}\rho_{10}, \quad (12e)$$

$$\dot{\rho}_{02} = i\Omega_1(t)\rho_{12} - i\Omega_2(t)(\rho_{00} - \rho_{22}) + i\omega_{20}\rho_{02} - \frac{1}{2}(\Gamma_{20} + \Gamma_{01})\rho_{02}, \quad (12f)$$

$$\dot{\rho}_{20} = -i\Omega_1(t)\rho_{21} + i\Omega_2(t)(\rho_{00} - \rho_{22}) - i\omega_{20}\rho_{20} - \frac{1}{2}(\Gamma_{20} + \Gamma_{01})\rho_{20}, \qquad (12g)$$

$$\dot{\rho}_{21} = -i\Omega_1(t)\rho_{20} + i\Omega_2(t)\rho_{01} - i\omega_{21}\rho_{21} - \frac{\Gamma_{20}}{2}\rho_{21}, \qquad (12h)$$

$$\dot{\rho}_{12} = i\Omega_1(t)\rho_{02} - i\Omega_2(t)\rho_{10} + i\omega_{21}\rho_{12} - \frac{\Gamma_{20}}{2}\rho_{12} \qquad (12k)$$

In Eqs. (12a) - (12k), $\rho_{00}$, $\rho_{11}$ and $\rho_{22}$ are the populations of levels 0, 1 and 2, respectively, $\rho_{01}$, $\rho_{02}$ and $\rho_{12}$ are the coherences between the energy levels.

## 3. The numerical evaluation of the population dynamics

In order to get the values of numerical integration of the above master equations (Eqs. (12a) - (12k)), we use the secant hyperbolic pulse laser, the pulse electric field of which can be written as

$$\boldsymbol{E}(t) = \boldsymbol{e}E_0 \, sec\, h[t/\tau_0]\cos\omega t, \qquad (13)$$

where $E_0$ is the constant amplitude of electric field, $\tau_0$ is the pulse width and $\omega$ is the laser field frequency. We can evaluate the value of the quantities as $\Omega_2(t) = (\boldsymbol{\mu}_{20}\cdot\boldsymbol{e})E(t)/\hbar$, $\Omega_1(t) = (\boldsymbol{\mu}_{01}\cdot\boldsymbol{e})E(t)/\hbar$ on condition that the dipole coupling coefficients $\mu_{20} = \boldsymbol{\mu}_{20}\cdot\boldsymbol{e}$, $\mu_{01} = \boldsymbol{\mu}_{01}\cdot\boldsymbol{e}$ and the pulse electric field $E(t)$ are given. In our numerical calculations, we choose the parameters as $\omega_{20} = 1.6\times 10^6\,\text{GHz}$, $\omega_{01} = 2.4\times 10^6\,\text{GHz}$, $\omega = 2\times 10^6\,\text{GHz}$, $\mu_{20} = \mu_{01} = 1.85\times 10^{-29}\,\text{Cm}$, thus $\Omega_1(t) = \Omega_2(t) = \Omega_0 \, sec\, h[t/\tau_0]\cos\omega t$, where $\Omega_0 = \frac{\mu\cdot E_0}{\hbar}$, which is dependent with the amplitude of electric field and the dipole moments, as well as the angle between the dipole moments and the direction of the electric field. At first, we set $\Gamma_{20} = \Gamma_{01} = 0$ for simplicity.

Figure 2 shows the population of three-level system of QD as function of time; we keep the pulse width constant while changing the Rabi frequency. Fig. 2(a) shows the complete depopulation of the excited state with the Rabi frequency $\Omega_0 = 0.6\times 10^6\,\text{GHz}$, where we set the pulse width $t_p = 10\,\text{ps}$. Fig. 2(b) shows that setting the Rabi frequency as $\Omega_0 = 0.530\times 10^6\,\text{GHz}$ results the superposition state with about 73% ground state and 26% excited state. One can see from Fig. 2(c) that equal occupation probabilities of the states are also possible by setting such suitable parameter as $\Omega_0 = 0.499 \times 10^6 \text{GHz}$, in which the state of the QD is the superposition of the 50% ground state and the 50% excited state. We can also found from Fig. 2(d) that population of the upper state can also be nearly 1, but not the same 1. These numerical calculations show that coherent control of the population transfer can be achieved by setting proper parameters such as the amplitudes of the electric field and the dipole coupling coefficients.

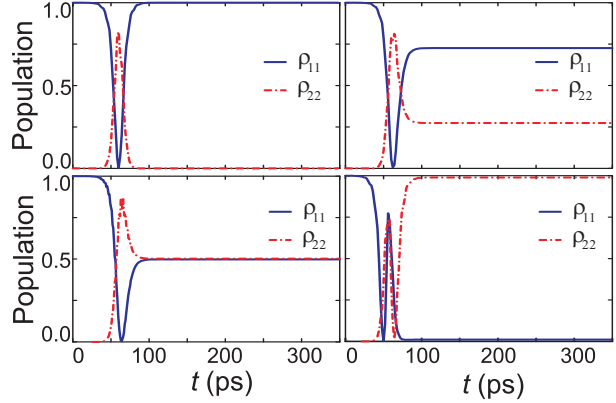

Fig. 2. Populations of QD versus time with different Rabi frequencies. (a) $\Omega_0 = 0.600 \times 10^6 \text{GHz}$, (b) $\Omega_0 = 0.530 \times 10^6 \text{GHz}$, (c) $\Omega_0 = 0.499 \times 10^6 \text{GHz}$, (d) $\Omega_0 = 0.785 \times 10^6 \text{GHz}$, where $t_p = 10\text{ps}$, $\Gamma_{20} = \Gamma_{01} = 0$.

We can find the dependency of the population with the Rabi frequencies of the system. Figure 3 shows the population as a function of Rabi frequency while keeping the pulse width constant as $t_p = 10\text{ps}$. One can see from the Fig. 3 easily the periodic dependency of the population with the Rabi frequency, however, the height of the peaks becomes lower as the Rabi frequency increases.

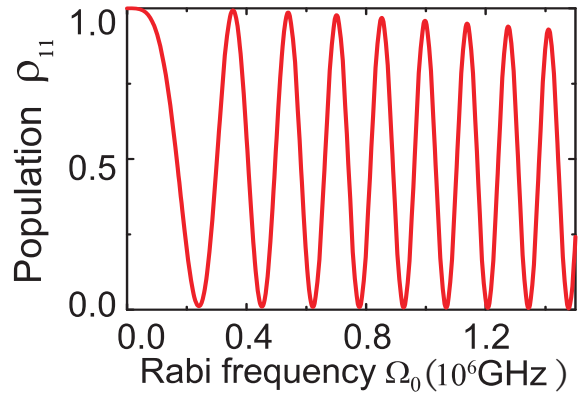

Fig. 3. Population as a function of Rabi frequency $\Omega_0$ ($t_p = 10$ps, $\Gamma_{20} = \Gamma_{01} = 0$).

Figure 4 shows the population of three-level system of QD as function of pulse width, where we set the Rabi frequency as $\Omega_0 = 0.4 \times 10^6$ GHz. From Fig. 4 one can see that the width of the population peaks is increased with increasing of the pulse width. We can also find that the maximum value of the peaks does not change monotonously.

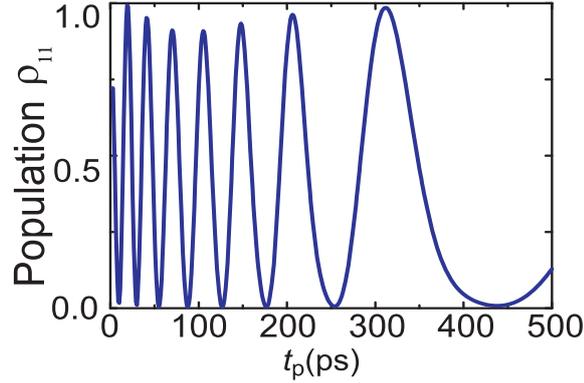

Fig. 4. Population as a function of pulse width $t_p$ ($\Omega_0 = 0.4 \times 10^6$ GHz, $\Gamma_{20} = \Gamma_{01} = 0$).

Finally, we can also consider the decoherence process of the QD system in the TPA process. Figure 5 shows the rate of the excitation state population to the ground state population versus decay rate of the states. Here, we set $\Gamma_{20} = \Gamma_{01} = \Gamma$ for simplicity. As we can see easily from Fig. 5, the rate of the populations of the QD system is decreased rapidly with increasing decay rate and saturated about from the decay rate 0.5ps$^{-1}$. This result gives an insight of the range of decay rate of the system considered.

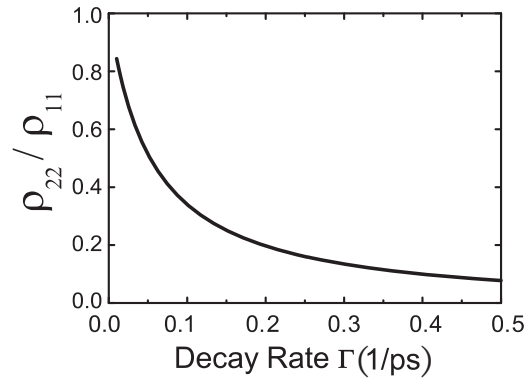

Fig. 5. Rate of Populations $\rho_{22}/\rho_{11}$ versus decay rate $\Gamma$.

The parameters used are $t_p = 10$ps, $\Omega_0 = 0.785 \times 10^6$ GHz.

## 4. Conclusion

In conclusion, we have investigated the influence of pulse width and Rabi frequency on the population dynamics of a three-level system in the two-photon absorption process. Our study gives valuable ranges of parameters such as pulse width, amplitude of the electric field and dipole coupling coefficients for practical applications. We confirmed that controlling pulse width and intensity of the electric field can be used to achieve population transfer on demand. The results can be extended to the case of more complex system such as, e. g., multilevel systems and they can be valuable for coherent quantum control schemes and quantum information processing.

## Acknowledgments

This work was supported by Key Project for Frontier Research on Quantum Information and Quantum Optics of Ministry of Education of D. P. R of Korea.